\documentclass[prb,twocolumn,showpacs,preprintnumbers,amsmath,amssymb]{revtex4}
\usepackage{graphicx}% Include figure files
\usepackage{dcolumn}% Align table columns on decimal point
\usepackage{bm}% bold math

\begin{document}

%\draft

\title{Evidence for a Novel State of Superconductivity in Noncentrosymmetric CePt$_{3}$Si : A $^{195}$Pt-NMR Study}

\author{M.~Yogi$^1$, Y.~Kitaoka$^1$, S.~Hashimoto$^2$, T.~Yasuda$^2$, R.~Settai$^2$, T.~D.~Matsuda$^3$, Y.~Haga$^3$, Y.~\={O}nuki$^{2,3}$, P.~Rogl$^4$ and E.~Bauer$^5$}

\address{$^1$Department of Materials Science and Technology, Graduate School of Engineering Science, Osaka University,  Osaka 560-8531, Japan}
\address{$^2$Graduate School of Science, Osaka University, Toyonaka, Osaka 560-0043, Japan}
\address{$^3$Advanced Science Research Center, Japan Atomic Energy Research Institute, Tokai, Ibaraki 319-1195, Japan}
\address{$^4$Institut f\"{u}r Physikalische Chemie, Universit\"{a}t Wien, A-1090 Wien, Austria}
\address{$^5$Institut f\"{u}r Festk\"{o}rperphysik, Technische Universit\"{a}t Wien, A-1040 Wien, Austria}
%\affiliation{%
%\affiliation{

\date{\today}
%\date{\today}

%\twocolumn[

%\maketitle

%\widetext

\begin{abstract}
We report on novel antiferromagnetic (AFM) and superconducting (SC) properties of noncentrosymmetric CePt$_{3}$Si through measurements of the $^{195}$Pt nuclear spin-lattice relaxation rate $1/T_1$.
In the normal state, the temperature ($T$) dependence of $1/T_{1}$ unraveled the existence of low-lying levels in crystal-electric-field multiplets and the formation of a heavy fermion (HF) state.
The coexistence of AFM and SC phases, that emerge at $T_{\rm N}=2.2$ K and $T_{\rm c}=0.75$ K, respectively,  takes place on a microscopic level.
CePt$_3$Si is the first HF superconductor that reveals a peak in $1/T_1$ just below $T_{\rm c}$ and, additionally, does not follow the $T^3$ law that used to be reported for most unconventional HF superconductors. We remark that this unexpected SC characteristics may be related with the lack of an inversion center in its crystal structure. 
\end{abstract}

\vspace*{5mm}

%]
\maketitle
%\narrowtext

%\section{Introduction}

In almost all previous studies on superconductors, it was assumed that the crystal has an inversion center, which makes it possible to 
separately consider the even (spin-singlet) and odd (spin-triplet) components of the superconducting (SC) order parameter (OP)  \cite{CePtSi_Anderson1,CePtSi_Anderson2}.
For example, Ce-based HF superconductors CeCu$_{2}$Si$_{2}$ and CeMIn$_{5}$ (M = Co, Rh, Ir) have a centrosymmetry in the crystal structure.
Although this is the case in most superconductors, there are some exceptions.
Very recently, Bauer {\it et al.} discovered a new heavy-fermion (HF) compound CePt$_{3}$Si that does not have centrosymmetry or an inversion center \cite{CePtSi_Bauer}.
The crystal structure of CePt$_{3}$Si belongs to the space group $P4mm$ (No.99), being isostructural with the ternary boride compound CePt$_{3}$B as shown in the inset of Fig.1 \cite{CePtSi_Sologub,CePtSi_Bauer}.
CePt$_{3}$Si exhibits antiferromagnetic (AFM) order below $T_{\rm N}=2.2$ K, and undergoes a SC transition at $T_{\rm c}=0.75$ K that was probed by the measurements of electrical resistivity $R(T)$ and specific heat $C_{\rm p}$.
As for the AFM ordering, $R(T)$ does not show a pronounced anomaly at $T = T_{\rm N}$, but $C_{\rm p}/T$ exhibits there a distinct peak \cite{CePtSi_Bauer}.
The entropy release at $T_{\rm N}$ is estimated to be as small as $\Delta S=0.22R\log2$.
This result refers to the development of Kondo-like interaction between $4f$ electrons and conduction electrons below 10 K that is evidenced from an almost logarithmic increase of $C_{\rm p}/T$ below 10 K as well.
With respect to the SC characteristics, only a small jump of $C_{\rm p}/T$ at $T_{\rm c}$ was found ($\Delta C_{\rm p}/\gamma T_{\rm c} \sim 0.25$), but an enhanced value of Sommerfeld coefficient $\gamma\sim 390$ mJ/mol K$^{2}$ suggests that HF superconductivity is realized in this compound.
Furthermore, $H_{\rm c2}$ at low $T$ exceeds the usual Pauli paramagnetic limiting field, indicative of a possibility of spin-triplet pairing.
But from a general argument that the lack of an inversion center does not always allow stable spin-triplet pairing, a mixing of spin-singlet and spin-triplet pairing was proposed for CePt$_{3}$Si \cite{CePtSi_Bauer}.

Very recently, neutron scattering experiments elucidated magnetic properties on CePt$_{3}$Si \cite{CePtSi_Metoki}.
An AFM Bragg reflection  was deduced at a wave vector ${\bf Q}=$(0,0,1/2) and (1,0,1/2), indicating that  Ce-$4f$ derived magnetic moments $\sim$0.3$\mu_{\rm B}$/Ce at $T=1.8$ K align ferromagnetically in the basal plane and stacked antiferromagnetically along the $c$-axis.
Its size is much smaller than the value expected from a Ce$^{3+}$ crystal electric field (CEF) ground state with either a $\Gamma_{7}$ or $\Gamma_{6}$ Kramers doublet.
This suggests that  Kondo-like interactions become effective, reducing the size of the ordered moments with respect to the simple localized picture. 
Consistency is found  with the results of specific-heat measurements that indicate only a small entropy release at $T_{\rm N}$ as mentioned above \cite{CePtSi_Bauer}.
The neutron scattering experiments also provided information for a CEF energy scheme \cite{CePtSi_Metoki}.
The most  remarkable finding is that  well defined CEF excitations were observed in this compound, although 4$f$ electrons are hybridized with conduction electrons to form heavy quasi-particles.
A similar observation was reported for the first Pr-$4f^2$-derived HF superconductor PrOs$_{4}$Sb$_{12}$ \cite{CePtSi_Kohgi_PrOs}.
In this case, it was revealed from band structure calculations and de Haas-van Alphen effect measurements that the 4$f$ electrons are almost localized \cite{CePtSi_Sugawara_PrOs}.

In order to gain insight into the possible OP symmetry of noncentrosymmetric CePt$_{3}$Si, several theoretical works have been put forward so far.
P. Frigeri $et\ al.$ have shown that in contrast to common believe, spin-triplet pairing is not entirely excluded in such systems \cite{CePtSi_Frigeri}.
Furthermore, the Pauli paramagnetic limit in $H_{\rm c2}(0)$ was analyzed for both spin-singlet and spin-triplet pairing.
CePt$_{3}$Si has the possibility of a $p$-wave spin-triplet pairing state ($\bm{d}(\bm{k}) = \hat{\bm{x}}k_{y} - \hat{\bm{y}}k_{x}$), that can explain the absence of the Pauli paramagnetic limit reported by Bauer $et\ al$.
On the other hand, noting that its superconductivity emerges under the background of AFM ordering, it seems more natural to assume a spin-singlet type of pairing, and Frigeri {\em et al.} argued that the Pauli paramagnetic limit is rendered less effectively by the presence of spin-orbit coupling arising from the broken inversion symmetry.
Band structure calculations revealed that a possible gap structure of CePt$_{3}$Si depends on the dimensionality of the SC OP, and if the SC OP corresponds to a one-dimensional representation, then the gap has line nodes where the Fermi surface crosses the high-symmetry planes or the boundaries of the Brillouin zone \cite{CePtSi_Samokhin}.

In this letter, we report on a microscopic study of CePt$_{3}$Si via measurements of nuclear magnetic resonance (NMR) of $^{195}$Pt.
Our aim is to clarify novel magnetic and SC characteristics inherent to CePt$_3$Si, the first HF superconductor without inversion symmetry, and provide microscopic evidence for the coexistence of antiferromagnetism and superconductivity.

%\section{Sample preparation}

We used polycrystalline and single crystals of CePt$_{3}$Si for the $^{195}$Pt-NMR measurement.
Polycrystalline samples were synthesized by high frequency melting.
Single crystals were successfully prepared by both the pulling method in a tetra-arc furnace and mineralization \cite{CePtSi_Metoki}.
For the $^{195}$Pt-NMR measurement, the samples are crushed into powder to make rf-field penetrate easily.

\begin{figure}[htbp]
  \begin{center}
   \includegraphics[keepaspectratio=true,height=45mm]{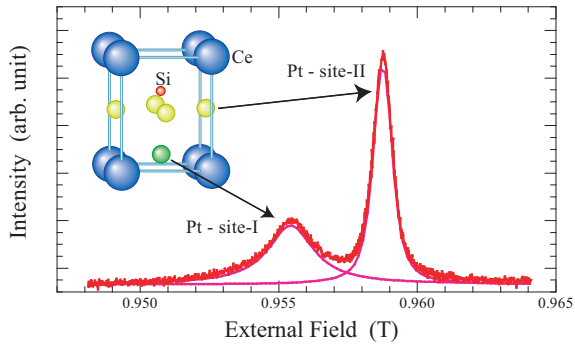}
  \end{center}
  \caption{(color online). The $^{195}$Pt-NMR spectrum for the oriented powder along the $c$-axis parallel to magnetic field ($H$) at 4.2 K and $f=8.9$ MHz ($H\sim$ 1 T).
}
\label{fig:spectra_4.2K.eps}
\end{figure}

Fig.1 displays the $^{195}$Pt-NMR spectrum at 8.9 MHz for  powder oriented along the $c$-axis parallel to $H$.
CePt$_{3}$Si has two inequivalent crystallographic Pt sites (see the inset of Fig.1 ).
One Pt site is surrounded by four Ce atoms within the $ab$ plane, labeled as site-I; the other one is labelled  as site-II.
The respective full-width-half-maximum (FWHM) in the Pt-NMR spectral shape are as small as 25 and 8 Oe at the site-I and site-II, assuring that the samples are well characterized.

%\section{Results and Discussions}
%\subsection{The Pt-NMR study}
%\subsubsection{NMR spectrum}

In the paramagnetic (PM) state above $T_{\rm N} = 2.2$ K, the Pt-NMR spectra at the site-I and -II exhibit positive values of Knight shift
but no significant changes.
Below $T_{\rm N}$, however, their FWHM's actually start to  increase, associated with the appearance of an internal field due to the onset of AFM order.
As a result, the spectra from the site-I and -II overlap below $T_{\rm N}$.
Note that the peak in the spectrum below $T_{\rm N}$ is mainly attributed to spectral weight from  the site-II.

%\subsubsection{Nuclear spin-lattice relaxation time $T_1$}

Since the $^{195}$Pt nuclei has a nuclear spin $I=1/2$, it follows a simple exponential form given by $\frac{M(\infty ) - M(t)}{M(\infty )} = \exp (-t/T_{1})$ where $M(\infty)$ and $M(t)$ are the nuclear magnetization for the thermal equilibrium condition and at a time $t$ after the saturation pulse, respectively.
$1/T_{1}$ is uniquely determined by a single component above $T_{\rm c}$.
At temperatures well below $T_{\rm c}$, however, two components  appear in $T_1$: a long component is attributed to an intrinsic relaxation, whereas the short one arises from the presence of vortex core induced by applying $H$.

%\paragraph{Normal State\\}

\begin{figure}[htbp]
  \begin{center}
    \includegraphics[keepaspectratio=true,height=70mm]{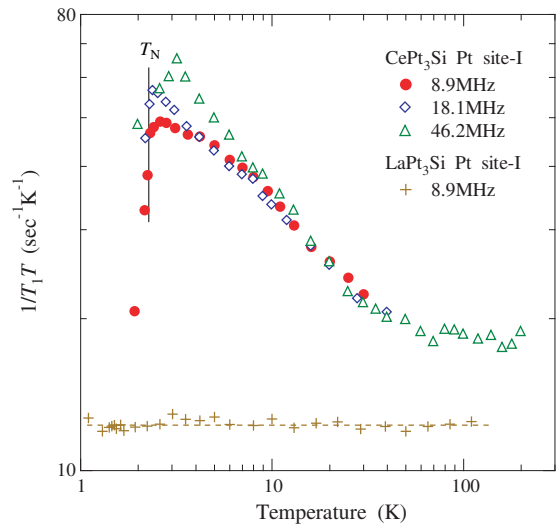}
  \end{center}
  \caption{(color online). The $T$ dependence of $1/T_{1}T$ at the site-I at $f=8.9$ (circle) MHz, 18.1 MHz (square), and 46.2 MHz (triangle) for CePt$_{3}$Si together with the result of LaPt$_{3}$Si (cross). A dashed line shows a Korringa law with $1/T_{1}T$ = 12.34 sec$^{-1}$K$^{-1}$ for LaPt$_{3}$Si in the measured $T$ range.}
  \label{fig:Fig2.eps}
\end{figure}

Fig.2 presents the $T$ dependence of $1/T_{1}T$ at the site-I at $f=8.9$ MHz, 18.1 MHz, and 46.2 MHz together with results of LaPt$_{3}$Si that are consistent with a Korringa law with $1/T_{1}T$ = 12.34 sec$^{-1}$K$^{-1}$ in the $T$ range measured.
Apparently, $1/T_{1}T$ in CePt$_{3}$Si is enhanced upon cooling due to the development of $4f$ derived magnetic fluctuations.
For temperatures above 10~K, $1/T_{1}T$ appears to be field independent, whereas below this characteristic temperature a significant 
$H$-dependence is deduced. This temperature coincides roughly with the CEF level energy  $\Delta E_1\sim$ 16 K.
Correspondingly to this, the $1/T_{1}T$ at $f=8.9$ MHz seems to  saturate below 10 K, whereas at $f=46.2$ MHz, $1/T_{1}T$ exhibits a sharp cusp at $T\sim 3$ K.  
These results suggest that the low-lying CEF level makes the relaxation behavior at low $T$ dependent on $H$.
Furthermore, the fact that $1/T_{1}T$ starts to decrease rapidly below $T_{\rm N}$ assures the onset of AFM order, consistent with the specific-heat result.
Unfortunately since the broadening in the spectrum below $T_{\rm N}$ makes it difficult to separately measure the spectrum at the site-I and II, $1/T_{1}$ cannot be measured below $T_{\rm N}$ at the site-I.

\begin{figure}[htbp]
  \begin{center}
    \includegraphics[keepaspectratio=true,height=70mm]{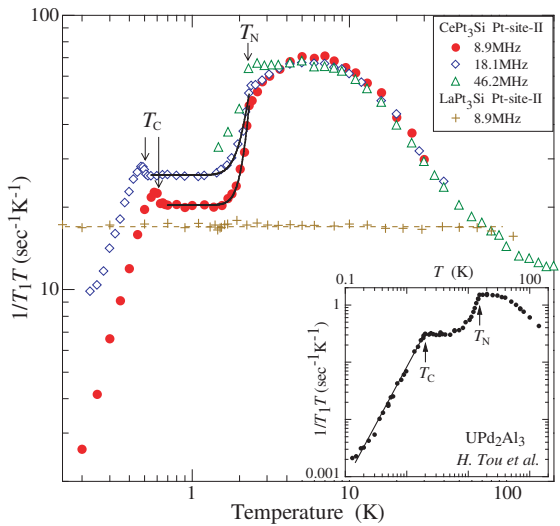}
  \end{center}
  \caption{(color online). The $T$ dependence of $1/T_{1}T$ at the site-II for CePt$_{3}$Si at $f$ = 8.9 MHz (circle), 18.1 MHz (square), and 46.2 MHz (triangle) together with the result of LaPt$_{3}$Si (cross). The dashed line shows a Korringa law with $1/T_{1}T$ = 17.03 sec$^{-1}$K$^{-1}$. The solid lines are least square fits to the data below $T_N$ using eq.(1). The inset presents the $T$ dependence of $1/T_{1}T$ of $^{27}$Al for the antiferromagnetic HF superconductor UPd$_{2}$Al$_{3}$ with $T_{\rm N}=14.5$ K and $T_{\rm c}=2$ K\cite{CePtSi_Tou}.}
 \label{fig:Fig3.eps}
\end{figure}

Instead, the $T$ dependencies of $1/T_{1}T$ at the site-II for CePt$_{3}$Si and LaPt$_{3}$Si are presented in a wide $T$ range in Fig.3 where the Korringa law $1/T_{1}T$ = 17.03 sec$^{-1}$K$^{-1}$ is valid at the site-II as well as at the site-I of LaPt$_{3}$Si. 
A difference in $1/T_{1}T$ is evident at the site-I and -II for CePt$_{3}$Si.
In contrast to the result at the site-I, the $1/T_{1}T$ at the site-II shows a shallow peak around 6 K at $H\sim 1$ T (8.9 MHz) and stays constantly below $\sim 8$ K at $H\sim 5$ T (46.2 MHz).
The latter result is consistent with the $4f$ derived HF state, because the value of $1/T_{1}T$ is more strongly enhanced than the value in LaPt$_{3}$Si as observed in either Ce or U-based HF compounds. 
This $H$ dependence of $1/T_{1}T$ observed at the site-II originates from the low-lying CEF level as was already observed for the site-I.
It is assumed that the reduction in $1/T_{1}T$ observed at low values of $H\sim$ 1 and 2 T is also caused by the low-lying CEF level.
This result contrasts with conventional Ce-based HF compounds where any well defined low-lying CEF level does not survive at low $T$. 
We like to point out that a similar $T$ dependence of $1/T_{1}T$ was recently reported for the Pr-based HF superconductor PrOs$_{4}$Sb$_{12}$, although the low-lying CEF level is a non-Kramers doublet, yielding a broad peak around 3.5 K \cite{CePtSi_Kotegawa}.

%\paragraph{AFM State\\}

The $T$ dependence of $1/T_{1}T$ undergoes a drastic decrease below $T_{\rm N}=2.2$ K without any critical divergence associated with the onset of AFM order as shown in Fig.3.
$1/T_{1}T$ below $T_{\rm N}$ is well reproduced by the following formula,
\begin{equation}
\frac{1}{T_{1}T} = A \exp\left( -\frac{\Delta}{k_{\rm B}T} \right) + C
\label{Korringa+gap}
\end{equation}
as indicated by the solid lines in Fig.3 for $f=8.9$ and 18.1MHz. 
Note that a $T_1T={\rm const}$ behavior is found for $T$ well below $T_{\rm N}$. 
The exponential decrease of $1/T_{1}T$ below $T_{\rm N}$ may be associated with a {\it gap} formed partially in the low-lying magnetic excitations spectrum.
The respective values of energy gap $\Delta/k_{\rm B}$ are estimated to be 23.3 K and 16.8 K at $f=8.9$ and 18.1 MHz.
The size of {\it gap} is reduced by applying magnetic fields.
The parameter $C$ of eq.(\ref{Korringa+gap}) corresponds to the quasi-particle contribution.
The respective values of $1/T_{1}T$ are about 20.3 sec$^{-1}$K$^{-1}$ and 26.2 sec $^{-1}$K$^{-1}$ at  $f=8.9$ and 18.1 MHz, being larger than $1/T_{1}T$ = 17.03 in LaPt$_{3}$Si.
This evidences that low-lying magnetic  excitations are gapped, but low-lying quasi-particle excitations are rather in a gapless regime, giving rise to the $T_1T={\rm const}$ law even under the background of AFM order. 
As the {\it gap} size is reduced with increasing $H$, the value of $1/T_{1}T$ increases, indicative of a $H$ induced transfer of low-energy spectral weight of quasi-particles.
The present experiment also clarifies the vital role of the $4f$ derived CEF effect in forming a HF state at low $T$ that is realized in virtue of the existence of a gap in the magnetic excitations in the AFM state.
Thus, the SC transition emerges behind this unique HF state which coexists with the AFM phase.  

%\paragraph{SC State\\}
In the SC state, the relaxation behavior of $\rm CePt_3Si$ is quite different from that observed in unconventional Ce or U-based HF superconductors reported so far (see Fig.3). 
Most HF superconductors display a $T^3$ power-law behavior that is consistent with a line-node gap below $T_{\rm c}$ without a coherence peak characteristic for conventional BCS superconductors \cite{CePtSi_Ishida_CCS,CePtSi_Kawasaki_CCS,CePtSi_Kohori_Rh,CePtSi_Mito_Rh,CePtSi_Kohori_CoIr,CePtSi_Zheng_Ir,CePtSi_Kawasaki_Co,CePtSi_Tou}.
As a typical example, the inset in Fig. 3 presents the $T$ dependence of $1/T_1T$ for the AFM HF superconductor UPd$_2$Al$_3$ that undergoes the AFM and SC transitions at $T_{\rm N}=14.5$ K and $T_{\rm c}=2$ K, respectively \cite{CePtSi_Tou}.
By contrast, it is unexpected that $1/T_1T$ in CePt$_3$Si exhibits a small peak just below $T_{\rm c}$. However, the observed peak in $1/T_1T$ is much smaller than that observed for conventional BCS superconductors.
In order to examine the $H$ dependence of the relaxation behavior below $T_{\rm c}$, the normalized value of $(1/T_{1}T)_{\rm SC}/(1/T_{1}T)_{\rm N}$ is plotted as a function of $T/T_{\rm c}$ at 8.9 MHz ($H\sim 1$ T) and 18.1 MHz ($H\sim 2$ T) as shown in Fig.4.
Here $(1/T_{1}T)_{\rm N}$ corresponds to the values at the normal state at 8.9 MHz ($H\sim 1$ T) and 18.1 MHz ($H\sim 2$ T).
Apparently, the peak in $(1/T_{1}T)_{\rm SC}/(1/T_{1}T)_{\rm N}$ is almost independent of $H$. $(1/T_{1}T)$ at $H\sim 2$ T seems to  saturate at low $T$.
Actually, the recovery curve of nuclear magnetization that depends on $H$ suggests that the relaxation process is primarily dominated by the presence of vortex cores introducing the normal-state region.
By contrast, $1/T_{1}T$ at 8.9 MHz ($H\sim 1$ T) continues to decrease down to $T$ =0.2 K, the lowest temperature measured.
Neither an exponential law nor $T^{3}$ behavior is observed as far as the data down to $T$ = 0.2 K are concerned. 
CePt$_{3}$Si is the first HF superconductor that exhibits a peak in $1/T_1T$ just below $T_{\rm c}$ and, moreover, does not follow the $T^3$ law reported for most of the unconventional HF superconductors.

\begin{figure}[htbp]
  \begin{center}
    \includegraphics[keepaspectratio=true,height=60mm]{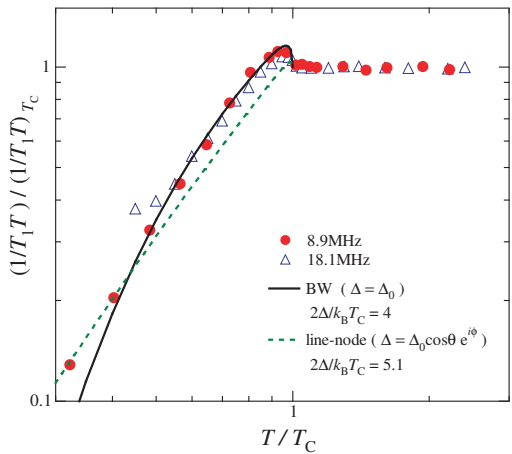}
  \end{center}
  \caption{(color online). The plot of $(1/T_{1}T)_{\rm SC}/(1/T_{1}T)_{\rm N}$ vs $T/T_{\rm c}$ at 8.9 MHz (circle) and 18.1 MHz (triangle). Here $(1/T_{1}T)_{\rm N}$ corresponds to the value at the normal state at 8.9 MHz ($H\sim 1$ T) and 18.1 MHz ($H\sim 2$ T). The solid line is a tentative fit calculated by applying the Balian-Werthamer model (BW isotropic triplet SC state) with a value of $2\Delta/k_{\rm B}T_{\rm c}=4$ \cite{CePtSi_BW}. Dashed line indicates a fit by a line-node gap model with $2\Delta/k_{\rm B}T_{\rm c}= 5.1$.}
\label{fig:Fig4.eps}
\end{figure}

In order to inspect the overall relaxation behavior of noncentrosymmetric CePt$_{3}$Si below $T_{\rm c}$, a tentative SC model was tried by applying the Balian-Werthamer model (BW isotropic spin-triplet SC state, solid line Fig.4) with a value of $2\Delta/k_{\rm B}T_{\rm c}=4$ \cite{CePtSi_BW}.
Note that the peak in $1/T_1T$ would indicate the presence of an isotropic energy gap, even though the {\it coherence effect}, inherent to the isotropic spin-singlet $s$-wave pairing state, is absent.
Moreover, $1/T_{1}T$ does not follow a simple power law behavior, indicating a line-node gap at the Fermi surface (dashed line, Fig. 4). 
It is interesting to note that the BW model describes fairly well the data just below $T_{\rm c}$. 
In order to conclude a more detailed structure of SC energy gap from the $1/T_1$ measurements, however, the experiment at further low temperatures and fields is required. 
In the new HF compound CePt$_{3}$Si, however, an inversion center is absent in its crystal symmetry.
Therefore, the novel relaxation behavior found below $T_{\rm c}$ may justify the possible existence of some kind of a new SC state being realized in the noncentrosymmetric CePt$_{3}$Si.

%\section{Conclusion}
In conclusion, the present $^{195}$Pt-NMR study on CePt$_3$Si has clarified novel electronic and magnetic properties in the PM and AFM state and a relevant new class of a HF SC state. 
The CEF energy separation of the low-lying level is as small as $\Delta E_1\sim 10-16$ K, consistent with neutron results. 
A HF state is realized for temperatures below $\sim$ 6 K. The magnetic excitation spectrum has a gap that depends on the magnetic field. 
The HF quasi-particles remain in a gapless regime in the $T$ region well below $T_{\rm N}$, giving rise to a $T_1T={\rm const}$ behavior.
The application of magnetic fields causes a reduction in size of magnetic gap and simultaneously the increase in spectral weight of quasi-particles at the Fermi level, having some relevance with the presence of low-lying CEF levels. The Pt-$T_1$ measurements have probed the  coexistence of AFM and SC orders on the microscopic level.
CePt$_3$Si is the first HF superconductor that reveals a peak in $1/T_1T$ just below $T_{\rm c}$ and additionally, does not follow the $T^3$ law, reported for most unconventional HF superconductors.
Thus, the novel relaxation behavior found  below $T_{\rm c}$ sheds light on the possibility of a new class of a SC state being realized in noncentrosymmetric CePt$_{3}$Si.
The experimental findings presented in this work deserve future theoretical studies to unravel the SC OP symmetry for the noncentrosymmetric compounds in general.

One of authors (M. Y. ) would like to thank G. -q. Zheng, and K. Ishida for valuable comments.
This work was partially supported by Grant-in-Aid for Creative Scientific Research (15GS0123), MEXT and the 21st Century COE Program by Japan Society for the Promotion of Science, and also by the Austrian FWF, P16370.

\end{document}